\documentclass[twocolumn,prb,amsmath,amssymb,amsfonts,superscriptaddress,floatfix,aps,showpacs,notitlepage]{revtex4-1}

\usepackage{graphicx}
\usepackage{epstopdf}
\usepackage{dcolumn}
\usepackage{bm}
\usepackage{rotating} 
\usepackage{color}
\usepackage{subfigure}
\usepackage{natbib}
\usepackage{enumitem}
\usepackage[Symbol]{upgreek}
\usepackage[symbol*]{footmisc}
\usepackage{lipsum}
\usepackage{amsmath}                    
\usepackage{amssymb}                    

\usepackage{textcomp}					

\begin{document}
\newcommand{\oscar}[1]{\textcolor{red} {#1}}
\renewcommand{\thetable}{\arabic{table}}

\title{Correlation effects and orbital magnetism of Co clusters}
\date{\today}
\author{L. Peters}
\email{L.Peters@science.ru.nl}
\affiliation{
Institute for Molecules and Materials, Radboud University Nijmegen, NL-6525 AJ Nijmegen, The Netherlands
 }
\author{I. {Di Marco}}
\affiliation{
Department of Physics and Astronomy, Uppsala University,
Box 516, SE-75120, Uppsala, Sweden
 }
\author{O. {Gr\aa n\"as}}
\affiliation{
Department of Physics and Astronomy, Uppsala University,
Box 516, SE-75120, Uppsala, Sweden
 }
\author{E. \c{S}a\c{s}{\i}o\u{g}lu}
\affiliation{
Peter Gr\"unberg Institut and Institute for Advanced Simulation, Forschungszentrum J\"ulich and JARA, 52425 J\"ulich, Germany
}
\author{A. Altun}
\affiliation{
Department of Physics, Fatih University, 34500 Istanbul, Turkey
}
\author{S. Rossen}
\affiliation{
Institute for Molecules and Materials, Radboud University Nijmegen, NL-6525 AJ Nijmegen, The Netherlands
 }
\affiliation{
Peter Gr\"unberg Institut and Institute for Advanced Simulation, Forschungszentrum J\"ulich and JARA, 52425 J\"ulich, Germany
}
\author{C. Friedrich}
\affiliation{
Peter Gr\"unberg Institut and Institute for Advanced Simulation, Forschungszentrum J\"ulich and JARA, 52425 J\"ulich, Germany
}
\author{S. Bl\"ugel}
\affiliation{
Peter Gr\"unberg Institut and Institute for Advanced Simulation, Forschungszentrum J\"ulich and JARA, 52425 J\"ulich, Germany
}
\author{M. I. Katsnelson}
\affiliation{
Institute for Molecules and Materials, Radboud University Nijmegen, NL-6525 AJ Nijmegen, The Netherlands
 }
\author{A. Kirilyuk}
\affiliation{
Institute for Molecules and Materials, Radboud University Nijmegen, NL-6525 AJ Nijmegen, The Netherlands
 }
\author{O. Eriksson}
\affiliation{
Department of Physics and Astronomy, Uppsala University,
Box 516, SE-75120, Uppsala, Sweden
 }

\begin{abstract}
Recent experiments on isolated Co clusters have shown huge orbital magnetic moments in comparison with their bulk and surface counterparts. These clusters hence provide the unique possibility to study the evolution of the orbital magnetic moment with respect to the cluster size and how competing interactions contribute to the quenching of orbital magnetism. We investigate here different theoretical methods to calculate the spin and orbital moments of Co clusters, and assess the performances of the methods in comparison with experiments. It is shown that density functional theory in conventional local density or generalized gradient approximations, or even with a hybrid functional, severely underestimates the orbital moment. As natural extensions/corrections we considered the orbital polarization correction, the LDA+U approximation as well as the LDA+DMFT method.  Our theory shows that of the considered methods, only the LDA+DMFT method provides orbital moments in agreement with experiment, thus emphasizing the importance of dynamic correlations effects for determining fundamental magnetic properties of magnets in the nano-size regime. 
\end{abstract}
\maketitle
\noindent

\section{Introduction}
The orbital magnetic moments of transition-metal bulk magnets are largely quenched~\cite{stohr-sigman}, while transition-metal surfaces have extremely large orbital moments~\cite{arvanitis,durr,eriksson}. Understanding the nature of the orbital moment is a problem of fundamental interest, and has over the years attracted much experimental attention through techniques such as XMCD~\cite {stohr} and ferromagnetic resonance~\cite{baberschke}. The orbital moment is important for several reasons. First, it contributes to the total magnetic moment of a system, and second it is together with the spin magnetic moment a measure of the extent of spin-orbit coupling in general and magnetic anisotropy in particular~\cite{vanvleck}. The latter property is also known to couple to the anisotropy of the orbital moment~\cite{bruno}. As a result, a detailed knowledge of orbital magnetism is crucial in designing new materials with desired hardness and saturation moments. Most interestingly, recent XMCD experiments~\cite{pered,lau} on transition-metal clusters showed huge orbital moments in comparison with their bulk counterparts. Hence these systems may possess a large magnetic anisotropy energy, which makes them potentially interesting for several technological applications. In this letter we investigate this possibility by means of several computational approaches.
%

Since the orbital magnetic moments of transition-metal clusters lie between those of the quenched values of the bulk systems and the large values of the isolated atoms, clusters provide a unique opportunity for studying the mechanisms that affect the orbital magnetic moments systematically. Unfortunately, there is up to now no theory available that can reproduce the experimentally observed large orbital moments for the transition-metal clusters. We formulate here, using several computational methods, a theory of orbital and spin magnetism for these clusters. In particular, we take Co clusters as a test case, because Co atoms possess the largest orbital moments among the transition-metals in all their forms, as clusters~\cite{pered,lau}, surfaces~\cite{eriksson} and bulk~\cite{stohr-sigman}.

Previous theoretical studies on the magnetic structure of transition-metal clusters~\cite{datta,agui1,agui2,lopez,gutsev,ma,guevara,co2dft5,fritsch,strand,strand2} focus primarily on the spin moment, due to the difficulties in estimating the orbital moment. The only theoretical study that addresses the orbital magnetic moment of pure clusters is focused on Co$_2$, which is technically more treatable. In Ref.~\onlinecite{lau} it is shown that the calculated spin moment in general is in reasonable agreement with experiment. Although the calculated spin and orbital magnetic moments of Co$_{13}$ clusters capped with Pt~\cite{co13pt} are available, there are no experimental data to compare with these calculations. The latter are also performed within GGA, which is in the present work shown to be inadequate for these systems, e.g. by severely underestimating the orbital magnetic moments with respect to experiment~\cite{pered,lau}. Therefore, it is expected that also the previous study on the orbital magnetism of Pt capped Co$_{13}$ clusters suffers from the inadequacy of GGA functionals.

The problem to face when calculating the orbital moment from conventional density functionals is the absence of Hund's second rule, which is the primary reason for the orbital moment and is driven by intra-shell electrostatic interaction. The crystal field effect competes against this interaction, and results in a quenching of the orbital magnetism. It will be shown here that for complex systems like clusters, a high level theory is required to properly describe the subtle competition among these effects. More precisely, it will be shown that plain density-functional theory (DFT)~\cite{dft1,dft2} in its conventional LDA/GGA~\cite{lda1,lda2,gga} or hybrid~\cite{hybrid} forms severely underestimates the orbital moment of Co clusters. Also an approximate consideration of Hund's second rule within extensions of plain DFT like 
the orbital polarization correction~\cite{rsptbook,olle} or the LDA$+U$ approach~\cite{ldauan,ldaulich} results in a severe underestimation. Among the theories explored in this study, the only one that is consistent with the measured orbital moments is the combination of DFT and dynamical mean-field theory (usually addressed as LDA+DMFT~\cite{dmft1,dmft2}), which treats onsite correlations and thus Hund's second rule exactly. This demonstrates the importance of dynamical correlations on orbital magnetism of magnetic transition-metal clusters as well as the fact that DMFT can give reasonable results for clusters too. 

For Co impurities in gold it has already been demonstrated that LDA+DMFT is required to produce the large orbital moments observed in experiments~\cite{chadov}. In this study, the authors used the spin-polarized T-matrix fluctuation exchange solver, because the correlation effects are not very strong and can be treated perturbatively. As a matter of fact, we did try to use this solver for the presently studied systems but found that it was inappropriate to describe the formation of orbital moments which are close to their atomic values. Therefore, we exploited the exact diagonalization routine for the impurity part of DMFT. 

\section{Theory} 
 \subsection{Theoretical methods}
 The focus of this work is on the calculation of the spin and orbital moments of pure Co clusters. For this purpose several codes based on density-functional theory (DFT), and extensions, have been used. The extensions are used to incorporate step by step a more sophisticated treatment of the onsite Coulomb repulsion. Namely it is this onsite Coulomb interaction that leads to the formation of local spin and orbital moments, which in atoms is summarized by the first and second Hund's rules. On the other hand the crystal field effects compete against this mechanism quenching both moments. A proper theory should take these two competing effects accurately into account, but at the moment such a theory exists only for pure atoms or dimers at most. Therefore we have explored several suitable techniques for our purposes. 
 
 The first method considered here is plain Kohn-Sham DFT, with an exchange-correlation functional in the local density approximation (LDA) as formulated by Perdew and Wang (PW)~\cite{lda1,lda2}, in the generalized-gradient approximation (GGA) as formulated by Perdew, Burke and Ernzerhof (PBE)~\cite{gga} and in the B3LYP hybrid approximation~\cite{hybrid}. In all these approximations the second Hund's rule is completely neglected, and orbital moments are induced by the spin moment through the spin-orbit coupling. Further, LDA and GGA are derived in the limit of a (nearly) uniform electron gas, while the hybrid functional treats the electron exchange of the inhomogeneous system partially exactly. Therefore, DFT in these forms only describes onsite Coulomb effects in a very rough approximation as far as orbital magnetism is concerned.
 
 The situation improves when an explicit onsite Coulomb repulsion term is considered, leading to a generalized Hubbard model~\cite{alexei}. The idea behind this is to combine DFT and the Hubbard model. Here, we exploit the fact that DFT works well for the (weakly correlated) delocalized electrons in the system, while the onsite Coulomb repulsion term is crucial for the description of (strongly correlated) localized electrons as known from studies of the Hubbard model. There are basically two methods available to approximately solve this generalized Hubbard model. The first is the static mean field approximation, i.e. the LDA+U method~\cite{ldauan,ldaulich}. This should describe to a certain extent the effects due to the second Hund's rule, although at the price of a forced broken symmetry, which is not a problem in the present case~\cite{held}. The second approach to this problem is based on the dynamical mean-field approximation~\cite{dmft1}, which leads to the LDA+DMFT approach~\cite{dmft2,held}. The LDA+DMFT approach becomes exact in the atomic limit or equivalently when hybridization effects can be neglected, in the non-interacting limit, and in the limit of an infinite number of nearest neighbors. Within this respect the regime of small clusters is rather far from the limit of infinite neighbors, although it has been shown that in practical terms this limit is reached very fast, even for a small number of nearest neighbors~\cite{dmft1,dmft2}. In order to evaluate the influence of hybridization effects we perform two types of LDA+DMFT calculations. The first is a simplified version of the LDA+DMFT method in the limit of zero hybridization, i.e. Hubbard-I approximation~\cite{hia1,hia2}. The second is a more accurate version, where the hybridization is considered within the exact diagonalization routine. 
 
Due to the inclusion of the onsite Coulomb interaction term, the Hubbard $U$ and Hund exchange $J$ parameters of Co clusters are required as an input for the LDA+U and LDA+DMFT calculations. It is not clear from the beginning what the Hubbard $U$ value in the cluster regime is but it is reasonable to assume that it is intermediate between the bulk value of about 3~eV and the atomic value of about 14~eV~\cite{chadov,ersoy,zaan}. To obtain the Hubbard $U$ and Hund exchange $J$ of Co clusters we performed calculations using the constrained random phase approximation (cRPA)~\cite{ersoy,arya}. The results are reported and discussed below.  
 
 
 In the LDA+U calculations there is a great risk to obtain a solution that corresponds to a local minimum instead of the global one. To avoid this problem we have used the method of Ref.~\onlinecite{ramp}, which consists in starting from a converged DFT calculation and then increasing $U$ and $J$ step-by-step. For completeness this type of calculation is compared with a LDA+U calculation starting from a converged DFT calculation, but without a step wise increase of the Hubbard $U$ and Hund exchange $J$ value. 
 
 Finally, most codes evaluate the orbital moments only within certain spheres around the atomic sites~\cite{olle,rsptbook}. However, here we have also evaluated the contribution to the orbital moment given by the interstitial region in between these spheres via the Modern Theory of Orbital Polarization~\cite{ceresolf,ceresol}.  
 
 Unfortunately, the plethora of these calculations could not be made by means of a single code. Therefore, different codes have been used for different purposes. Calculations based on LSDA and GGA were made using \textit{RSPt}~\cite{rsptbook}, \textit{VASP}~\cite{vasp} and \textit{Quantum ESPRESSO}~\cite{espresso} (with orbital moment obtained from Modern Theory of Orbital Polarization~\cite{ceresolf,ceresol}). The cRPA calculations of Hubbard $U$ and Hund exchange $J$ were made using \textit{FLEUR}~\cite{fleur} and \textit{SPEX}~\cite{spex}. The LSDA+U calculations were made with \textit{VASP}, whereas the LSDA+DMFT calculations were made with \textit{RSPt}. All the computational details are contained in the Appendix.

 \subsection{Hubbard $U$ and Hund exchange $J$ parameters}
 The Hubbard $U$ and Hund exchange $J$ parameters are required as an input for LDA+U and LDA+DMFT calculations. The cRPA method was used to calculate these parameters, which are reported in Table~\ref{hubujrpa} for Co$_{2}$ to Co$_{7}$. We find that $U$ and $J$ are slightly different for inequivalent atomic sites in a given cluster. Therefore, the values shown in Table~\ref{hubujrpa} are average values.

 \begin{table}[b]
 	\begin{center}
 		\begin{tabular}{|l|c|c|c|c|c|c|}
 			\hline
 			\textbf{cRPA} & $Co_{2}$ & $Co_{3}$ & $Co_{4}$ & $Co_{5}$ & $Co_{6}$ & $Co_{7}$ \\
 			\hline
 			$U$ (eV) &  9.7 & 8.8 & 8.3 & 7.7 & 7.3 & 7.2 \\
 			\hline
 			$J$ (eV) & 0.8 & 0.7 & 0.7 & 0.7 & 0.7 & 0.7\\
 			\hline
 		\end{tabular}
 		\caption{The Hubbard $U$ and Hund exchange $J$ parameters in eV obtained from cRPA calculations for Co$_{2}$ to Co$_{7}$ clusters.}
 		\label{hubujrpa}
 	\end{center}
 \end{table}  
 
 From Table~\ref{hubujrpa} one can observe that the Hubbard $U$ value decreases with increasing size, which indicates that the screening becomes more effective with increasing cluster size. Comparing these results with the $U$ and $J$ values predicted for Co bulk, $U=2-3$~eV and $J=0.7-0.9$~eV, it appears that for the clusters sizes considered the $U$ value is significantly larger, but $J$ is about the same~\cite{chadov,ersoy}. Note that it is well known that the Hund exchange $J$ is an atomic like quantity which is practically system independent. Therefore, it is not unexpected to find the Hund exchange $J$ to be independent of cluster size and almost equal to the bulk value.
 
We note here that in the calculation of the Hubbard $U$ and Hund exchange $J$ parameters within the cRPA method the $d$-$d$ screening channel is excluded. Therefore, the Hubbard $U$ and Hund exchange $J$ values of Table~\ref{hubujrpa} can be directly used for a LDA+DMFT calculation, where the $d$-$d$ screening is taken into account explicitly. As explained in the main text, to allow a better comparison among the LDA+DMFT calculations for different cluster sizes, the same value of 8~eV was used for $U$.



\section{Results}

\subsection{Spin and orbital moments from GGA and LDA+U}
We start by reporting on the spin and orbital moments from GGA (PBE)~\cite{gga} and LDA+U in Table~\ref{vaspso1}. These calculations were made using the VASP code~\cite{vasp}, and were focused on clusters of different size, from Co$_{2}$ to Co$_{9}$.  To allow a better comparison, also for the LDA+U method the GGA (PBE) functional was used. Thus, note that the nomenclature LDA+U in this work should be interpreted as GGA+U. Further, in order to analyze the $U$ dependence of the spin and orbital moment, two sets of LDA+U calculations were performed. One set, labeled as LDA+U(1), is for a Hubbard $U$ corresponding to the bulk value of 3~eV. The other one, labeled as LDA+U(2), is instead for a $U$ calculated appropriately for the cluster regime (see Table~\ref{hubujrpa}), which is of 7~eV. For both calculations the same $J$ value of 0.9~eV is used, as in Ref.~\onlinecite{chadov}.  

\begin{table}[!htb]
\begin{center}
\begin{tabular}{|l|c|c|c|c|c|c|c|c|}
\hline
Method & $Co_{2}$ & $Co_{3}$ & $Co_{4}$ & $Co_{5}$ & $Co_{6}$ & $Co_{7}$ & $Co_{8}$ & $Co_{9}$ \\
\hline
GGA &  2.08 & 2.33 & 2.50 & 2.60 & 2.33 & 2.14 & 2.0 & 2.11 \\
        &  0.34 & 0.21 & 0.14 & 0.15 & 0.12 & 0.13 & 0.12 & 0.12 \\
\hline
LDA+U(1) & 1.99 & 2.32 & 2.49 & 2.59 & 2.32 & 2.13 & 1.99 & 2.10\\
                 & 0.28 & 0.30 & 0.31 & 0.29 & 0.32 & 0.30 & 0.30 & 0.26\\
\hline
LDA+U(2) & 1.98 & 2.32 & 1.97* & 2.00* & 2.32 & 2.13 & 1.94* & 1.98*\\
                 & 0.27 & 0.29 & 0.28 & 0.27 & 0.27 & 0.26 & 0.27 & 0.25\\
\hline
Experiment & -- & -- & -- & -- & -- & -- & 2.6 & 2.1 \\
                   & -- & -- & -- & -- & -- & -- & 0.7 & 0.65\\
\hline
\end{tabular}
\caption{The spin (upper line) and orbital (lower line) moments in \textmu$_\mathrm{B}$/atom obtained from GGA, LDA+U and experiment~\cite{pered}. Here LDA+U(1) and LDA+U(2) correspond respectively to a LDA+U calculation with $U=3$~eV and $J=0.9$~eV, and $U=7$~eV and $J=0.9$~eV. The asterisks symbol for the LDA+U(2) method indicates that instead of a fully ferromagnetic structure an antiferromagnetic structure is obtained as the ground state. Here antiferromagnetic means that at some site(s) the magnetic moment is pointing in the opposite direction with respect to the other sites. For the reader's convience the experimentally observed orbital moment for the isolated atom and (hcp) bulk are 3 (for a $d^{7}$ configuration) and 0.13~\textmu$_\mathrm{B}$/atom~\cite{orbexp}.}
\label{vaspso1}
\end{center}
\end{table} 

The comparison of the computational results with the available experimental values~\cite{pered,lau} reveals that both GGA and LDA+U severely underestimate the orbital moments, while the computed spin moments are quite close to the experimental values (Table~\ref{vaspso1}). Further, it can be observed that LDA+U calculations in general result  in an orbital moment which is roughly a factor 2 larger than the value of GGA. This increase can be understood from the fact that the GGA calculations do not give any account of the orbital polarization induced by the second Hund's rule, while this contribution is partially described in LDA+U~\cite{solov}. The comparison between LDA+U(1) and LDA+U(2) emphasizes that spin and orbital moments do not depend strongly on the Hubbard $U$ parameter nor the cluster size. However, it is important to mention that for the LDA+U(2) setup for some cluster sizes (i.e. Co$_{4}$, Co$_{5}$, Co$_{8}$ and Co$_{9}$) an antiferromagnetic magnetic structure was favoured with respect to the ferromagnetic structure. More precisely, for Co$_{4}$ a magnetic ground state with two moments pointing up and two moments pointing down was found. For Co$_{5}$ there were four moments pointing up and one down, for Co$_{8}$ six moments were pointing up and two down, and for Co$_{9}$ eight moments were pointing up and one moment was pointing down. For these antiferromagnetic structures the values in Table~\ref{vaspso1} correspond to the site averaged absolute value of the spin and orbital moment. 
The experimental data of Refs.~\onlinecite{pered,lau,jeroen} indicate a ferromagnetic alignment of the Co moments for Co$_{4}$, Co$_{5}$, Co$_{8}$ and Co$_{9}$, hence the results of LDA+U(2) in Table~\ref{vaspso1} are inconsistent with experiments.

It is interesting to analyze the GGA calculations for Co$_{2}$ more in detail. Namely this calculation can be compared with the work of Refs.~\onlinecite{co2dft5,fritsch}. For the ground state of Co$_{2}$  a theoretical orbital moment of 0.39~\textmu$_\mathrm{B}$/atom and a spin moment of 1.95~\textmu$_\mathrm{B}$/atom were reported in Ref.~\onlinecite{co2dft5}, while in Ref.~\onlinecite{fritsch} the values of orbital and spin moments were respectively 1~\textmu$_\mathrm{B}$/atom and 2.05~\textmu$_\mathrm{B}$/atom. By using different starting densities for our self-consistent calculations, we managed to obtain a state with an orbital moment of 0.34~\textmu$_\mathrm{B}$/atom and a spin moment of 2.08~\textmu$_\mathrm{B}$/atom (see GGA result in Table~\ref{vaspso1}) and another state with an orbital moment of 0.94~\textmu$_\mathrm{B}$/atom and a spin moment of 2.11~\textmu$_\mathrm{B}$/atom. The former state was found to be 31~meV lower in energy with respect to the latter. This shows that it is possible to stabilize different stable and meta-stable configurations, and may explain the different results of Refs.~\onlinecite{co2dft5,fritsch}.
The small discrepancies between our values for the magnetic moments and the values of Refs.~\onlinecite{co2dft5,fritsch} are reasonable in terms of slight changes in the computational strategies and in the exchange-correlation functionals.   
Therefore, the two states with different orbital moments that were reported earlier should be interpreted as two different energy minima, and the state with low orbital moment is the ground state of the GGA functional.

 
 \subsection{Co$_{4}$ as a test case}
 None of the GGA or LDA+U results in Table~\ref{vaspso1} reproduce the experimental orbital moment, and there could be several reasons for this discrepancy. For example the XMCD experiment is performed on charged clusters, while the theoretical calculations are for neutral clusters. Another reason could be the consideration of an erroneous geometry. For Co$_{4}$ it is for example known from indirect vibrational spectroscopy experiments~\cite{jeroen} that the geometry is a planar rhombus, while theory not always finds this to be lowest in energy~\cite{datta,agil}. Further, the employment of an inappropiate functional could also lead to a discrepancy. In order to test the influence of the ionization of the cluster (i.e. charge), geometry and functional, Co$_{4}$ is used as a test case. We selected Co$_{4}$ as a test case, since each atom has already a three-fold coordination while the computational effort is still manageable for exploring different methods. In Table~\ref{co4test} the spin and orbital moments of Co$_{4}$ are reported, as obtained via various approaches.
 
\begin{table}[!hb]
 	\begin{center}
 		\begin{tabular}{|l|c|c|}
 			\hline
 			Method & Spin moment & Orbital moment   \\
 			& (\textmu$_\mathrm{B}$/atom) & (\textmu$_\mathrm{B}$/atom)\\
 			\hline
 			LDA (planar) RSPt & 2.44  & 0.10  \\
 			\hline
 			LDA (tetra) RSPt  & 2.44   & 0.12   \\
 			\hline
 			GGA (planar) VASP & 2.50 & 0.17 \\
 			\hline
 			GGA (tetra) VASP  & 2.50 & 0.14 \\
 			\hline
 			B3LYP (planar) VASP & 2.50 & 0.25 \\
 			\hline
 			B3LYP (tetra) VASP   & 2.50 & 0.20  \\
 			\hline
 			GGA+OPC (planar) RSPt &  2.48 & 0.33  \\
 			\hline
 			GGA+OPC (tetra) RSPt   & 2.48 & 0.21  \\
 			\hline
 			LDA+U(1) (planar) VASP & 2.49 & 0.31 \\
 			\hline
 			LDA+U(1) (tetra) VASP   & 2.49 & 0.31  \\
 			\hline
 			LDA+U(2) (planar) VASP* &  1.96 & 0.26  \\
 			\hline
 			LDA+U(2) (tetra) VASP*   & 1.97 & 0.28  \\
 			\hline
 			GGA charged (planar) VASP & 2.25 & 0.18 \\
 			\hline
 			GGA charged (tetra) VASP   & 1.75 & 0.13   \\
 			\hline
 		\end{tabular}
 		\caption{The spin and orbital moments (in \textmu$_\mathrm{B}$/atom) of Co$_{4}$ as obtained from different methods are given. The geometry is indicated within the round brackets, where 'planar' refers to the planar rhombus and 'tetra' to the (distorted) tetrahedron. Further, OPC refers to the orbital polarization correction~\cite{olle}. The asterisks indicate that an antiferromagnetic ground state is obtained instead of a ferromagnetic one. }
 		\label{co4test}
 	\end{center}
 \end{table}
 
The analysis of the results of Table~\ref{co4test} is much simplified by first noticing that the theoretical data reported in Table~\ref{vaspso1} show that the orbital moment does not change much in the range of cluster sizes considered here. This is consistent with the experiments of Refs.~\onlinecite{pered,lau}, where the orbital moment is found to exhibit only a weak dependence on the cluster size. Taking theory and experimental results together, an orbital moment of at least about 0.7~\textmu$_\mathrm{B}$/atom is \textit{naively} expected for Co$_{4}$. Later it will be shown that this seems to be correct. One can immediately notice that none of the orbital moments reported in Table~\ref{co4test} is close to a value of 0.7~\textmu$_\mathrm{B}$/atom, reflecting a problem with the theoretical description. 

Further, the results in Table~\ref{co4test} show that the geometry hardly influences the values of the spin and orbital moments. Very small changes are also found when changing the exchange-correlation functional from LDA to GGA, while the hybrid (B3LYP) functional leads to a somewhat  larger increase in orbital magnetism. Considering a charged cluster leads instead to an interesting dependence of the magnetic moments on the assumed geometry. For the planar arrangement spin and orbital moments are similar to those of a non-charged cluster, while for the tetrahedron geometry the charge has a large influence on the spin moment. The orbital polarization correction increases the orbital moments obtained with LDA and GGA slightly and makes the results very similar to those obtained with a hybrid functional. The largest values for the orbital moment are obtained from the LDA+U calculations, albeit with values far from the expected experimental results. However, for the large $U$ setup, i.e. LDA+U(2), an unexpected antiferromagnetic ground state is obtained, which consists for both geometries in a configuration where the magnetic moments point up at two atomic sites and down at the other two. 

 Another possible source of error not considered so far could be the contribution of the interstitial region to the orbital moment. Namely in RSPt~\cite{rsptbook} and VASP only the contribution to the orbital moment within a certain sphere around the atomic sites is considered. Therefore, the Quantum Espresso code~\cite{espresso} was used in order to evaluate the interstitial region contribution to the orbital moment. For Co$_{3}$, Co$_{4}$ and Co$_{5}$ respectively the interstitial contribution to the total orbital moment was found to be 1~\%, 4~\% and 15~\%. Taking 15~\% of the largest value for the orbital moment found so far, i.e. 0.3~\textmu$_\mathrm{B}$/atom, gives roughly an 0.05~\textmu$_\mathrm{B}$/atom orbital moment contribution of the interstitial region. This is obviously much too small to cover the difference between experiment and theory. 

\subsection{LDA+DMFT}
From the results obtained for Co$_{4}$ one can conclude that for all the approaches tried, the orbital moment is underestimated with respect to our extrapolation of the experiments. Thus, replacing the exchange correlation functional with one of the most common formulations, adjusting the geometry and charge of the clusters, or including the interstitial contributions does not lead to a substantial increase in the orbital moments. Therefore, we resort here to a more sophisticated method, the LDA+DMFT approach, where atomic-like effects are treated via a multi-configurational solution of the many-body problem~\cite{dmft1,dmft2}. While in LDA+U it is a common practice to perform calculations for different values of the Hubbard $U$, in LDA+DMFT one can use directly the values calculated through constrained random-phase approximation (cRPA)~\cite{ersoy,arya}, which removes a parameter from the calculations and makes them fully ab-initio. These values are usually not used for LDA+U due to that this approach does not account for the dynamical screening due to 3d electrons themselves, which reduces the effective value of $U$ by an unknown amount~\cite{dmft2}. Our calculated values of the Hubbard parameter $U$ and Hund exchange parameter $J$ are reported in Table~\ref{hubujrpa}. For simplicity, and for offering a better comparison among clusters of different size, we used $U=8$ eV for all calculations. We also checked the effect of a larger $U=9$ eV for Co$_2$, in agreement with Table~\ref{hubujrpa}, and we found no major changes (see below). We consider two different approximations of the local impurity problem to investigate separately the influence of the static crystal field and hybridization (kinematic effect) on the orbital moment. First, we evaluate the performance of the LDA+DMFT method without including any effect of the hybridization, which corresponds to the Hubbard-I approximation~\cite{hia1,hia2}. Then, a more accurate solution is obtained by considering the hybridization effects through the exact diagonalization solver~\cite{patrick}. 

\begin{figure}[!ht]
\begin{center}
\includegraphics[trim=0 0 0 0, clip, width=9cm, scale=0.5]{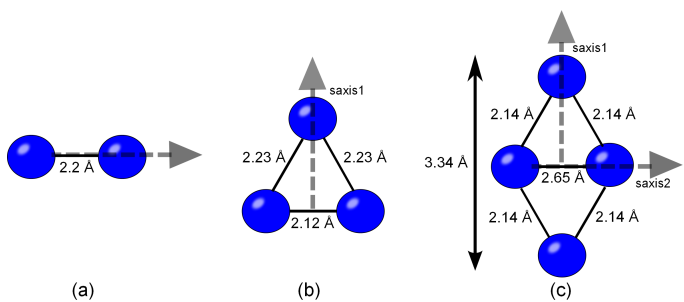}
\end{center}
\caption{The geometry and spin axes are indicated for (a) Co$_{2}$, (b) Co$_{3}$ and (c) planar rhombus Co$_{4}$. For Co$_{3}$ the second spin axis (saxis2) is orthogonal to the triangular plane.} 
\label{co234}
\end{figure}

The Hubbard-I approximation calculations were performed for clusters ranging from Co$_2$ to Co$_9$, while experimental data are available for only Co$_8$ and Co$_9$ (see Table 1)~\cite{pered,lau}. For all cluster sizes the same geometries as those in Table 1 are considered. The only exception is Co$_4$ for which a planar rhombus is considered, because it is experimentally known to be the ground state instead of the (distorted) tetrahedron~\cite{jeroen}.  Further, we analyse different directions for the magnetization for the Hubbard-I approximation calculations of Co$_2$, Co$_3$ and Co$_4$, as shown in Fig.~\ref{co234}. Since from Table~\ref{tabhia} one can infer that the direction of the magnetization axis is not so crucial for the magnetic properties, for clusters of larger size only one direction is reported. Co$_5$ is a trigonal bipyramid with the spin axis orthogonal to the common base of both pyramids. Co$_6$ is an octahedron, where the spin axis 'connects' the most distant atoms. Co$_7$ is a capped octahedron for which the spin axis is chosen to 'connect' the most distant atoms of the octahedron-part of the structure. Co$_8$ and Co$_9$ are respectively a bicapped and a distorted tricapped octahedron for which the spin axis is chosen equivalently to Co$_7$. 

In Table~\ref{tabhia} the spin and orbital moments obtained within the Hubbard-I approximation are shown. Since Hund's second rule effects might be sensitive to a change in the Hund's rule $J$ parameter, we performed calculations for $J=0.7$~eV and $J=0.5$~eV. The former calculations correspond to the first of the split 'Spin moment' and 'Orbital moment' columns, while the latter to the second. It appears that both spin and orbital moment are hardly influenced by this change of $J$. It is also clear that already within this simplified version of the LDA+DMFT method, the orbital moments of Co$_8$ and Co$_9$ are in very good agreement with the experimental values (see Table 1). Further, Table~\ref{tabhia} clearly shows that the orbital moment does almost not depend on the cluster size, which was also observed from Table 1 and experiments~\cite{pered,lau}. An exception here is Co$_2$, which has a substantially larger orbital moment than observed for larger clusters. As mentioned above for Co$_{4}$, judging the quality of the results for clusters smaller than Co$_8$ requires an extrapolation of the experimental data obtained for Co$_8$ and Co$_9$. This extrapolation is based on the experimental value for Co$_8$ and on the fact that both experiments and calculations (via DFT and LDA+U) show a very weak depedence of the orbital moment on the cluster size (see Table~\ref{vaspso1} and Refs.~\onlinecite{pered,lau}). Thus, from this extrapolation orbital moments of approximately 0.7~\textmu$_\mathrm{B}$/atom are expected for clusters from Co$_2$ to Co$_7$. The results of Table~\ref{tabhia} clearly confirm this expectation, with the notable exception of Co$_2$, which will be discussed in more detail below. 

\begin{table}[!hb]
\begin{center}
\begin{tabular}{|l*{4}{|c}|}
\hline
System & \multicolumn{2}{c|}{Spin moment} & \multicolumn{2}{c|}{Orbital moment}   \\
             & \multicolumn{2}{c|}{(\textmu$_\mathrm{B}$/atom)} &  \multicolumn{2}{c|}{(\textmu$_\mathrm{B}$/atom)} \\ 
\hline
Co$_{2}$ & { } 2.94 { }&2.95& { }{ } 1.29  { }{ }& 1.32  \\
\hline
Co$_{3}$ saxis1 & { }2.97 & 2.99 & { }{ } 0.75 { }{ }& 0.75  \\
\hline
Co$_{3}$ saxis2 & { }2.98 & - & { }{ } 0.73 { }{ }& - \\
\hline
Co$_{4}$ saxis1 & { }2.49 & - & { }{ } 0.64 { }{ }& - \\
\hline
Co$_{4}$ saxis2 & { }2.49 & 2.49 & { }{ } 0.74 { }{ }& 0.74 \\
\hline
Co$_{5}$ & { }2.58 & 2.58 & { }{ } 0.73 { }{ }& 0.74 \\
\hline
Co$_{6}$ & { }2.45 & 2.47 & { }{ } 0.69 { }{ }& 0.68 \\
\hline
Co$_{7}$ & { }2.41 & 2.42 & { }{ } 0.72 { }{ }& 0.73 \\
\hline
Co$_{8}$ & { }2.73 & 2.74 & { }{ } 0.67 { }{ }& 0.67  \\
\hline
Co$_{9}$ & { }2.74 & 2.76 & { }{ } 0.69 { }{ }& 0.69 \\
\hline
\end{tabular}
\caption{The spin and orbital moments in \textmu$_\mathrm{B}$/atom calculated with the LDA+DMFT method within the limit of zero hybridization, i.e. Hubbard-I approximation, for Co$_{2}$ to Co$_{9}$ clusters. Note that Co$_8$ and Co$_9$ are measured experimentally with an orbital moment of respectively 0.7 and 0.65~\textmu$_\mathrm{B}$/atom~\cite{pered,lau}.  Here, saxis refers to spin axis direction. For Co$_{3}$ saxis1 is in the triangular plane and saxis2 is orthogonal to the triangular plane (see Fig.~\ref{co234}). For Co$_{4}$ both spin axes are in plane (see Fig.~\ref{co234}). Further, the first column of the 'Spin moment' and 'Orbital moment' column refers to a $J=0.7$~eV calculation, while the second to that of $J=0.5$~eV. }
\label{tabhia}
\end{center}
\end{table}

Furthermore, it can be observed from Table~\ref{tabhia} that the spin moment of Co$_{8}$ is in good agreement with experiment, while for Co$_{9}$ it is a bit off. On the other hand the spin moment of Co$_{9}$ is in good agreement with that of Co$_{8}$ and smaller clusters, which is the trend one would expect for a ferromagnetically coupled system. Regarding such a trend, the spin moment of Co$_{9}$ is in very good agreement with what one would expect from the experimental data of Ref.~\onlinecite{lau}.

In the following we will report on how a more accurate version of LDA+DMFT, i.e. also including hybridization effects within the exact diagonalization solver, changes the orbital moments. From the very good match of the Hubbard-I results with experiment for Co$_8$ and Co$_9$, one would expect hybridization effects to be small. Due to computational reasons, and also in light of the observed size independence of the orbital moment, we have considered only Co$_{2}$, Co$_{3}$ and Co$_{4}$ clusters for these more accurate LDA+DMFT calculations. For these clusters the same geometries as for the Hubbard-I approximation calculations are used. In Fig.~\ref{co234} the used geometries together with the directions of the magnetization axis under consideration are depicted.

\begin{table}[!hbt]
\begin{center}
\begin{tabular}{|l|c|c|}
\hline
System & Spin moment & Orbital moment   \\
             & (\textmu$_\mathrm{B}$/atom) & (\textmu$_\mathrm{B}$/atom)\\
\hline
Co$_{2}$ & 2.97  & 0.72  \\
\hline
Co$_{2}$ $U=9$~eV  & 2.97   & 0.71   \\
\hline
Co$_{2}$ $IAD=2.4$~\AA & 2.97 & 0.71 \\
\hline
Co$_{3}$ saxis1   & 2.98 & 0.86 \\
\hline
Co$_{3}$ saxis2 & 2.98 & 0.76 \\
\hline
Co$_{4}$ saxis1   & 2.47 & 0.73  \\
\hline
Co$_{4}$ saxis2 & 2.48  & 0.80  \\
\hline
\end{tabular}
\caption{The spin and orbital moments in \textmu$_\mathrm{B}$/atom calculated with the LDA+DMFT method are printed for Co$_{2}$, Co$_{3}$ and Co$_{4}$ clusters. Here IAD stands for interatomic distance, which is 2.2~\AA~for the Co$_{2}$ calculations without IAD specification. Further, saxis refers to spin axis direction. For Co$_{3}$ saxis1 is in the triangular plane and saxis2 is orthogonal to the triangular plane (see Fig.~\ref{co234}). For Co$_{4}$ both spin axes are in plane (see Fig.~\ref{co234}). }
\label{tabdmft}
\end{center}
\end{table}

In Table~\ref{tabdmft} the spin and orbital moments obtained within the more accurate execution of the LDA+DMFT method are shown. From this table it is observed that the effect of the hybridization on the spin and orbital moments is indeed small for Co$_3$ and Co$_4$, while it is large for Co$_2$. This large influence on the orbital moment for Co$_2$ can be traced back to the energy difference between the many body eigenstates obtained in the Hubbard-I approximation. Namely, for Co$_2$ the energy difference between the ground state and the first two higher lying states is at least an order of magnitude smaller than what is observed for Co$_3$ and Co$_4$. Since for clusters from Co$_5$ to Co$_9$ this energy difference is of the same order of what is found for Co$_3$ and Co$_4$, hybridization effects should be small also for these clusters. This discussion leads us to conclude that our calculated value of the orbital moment for clusters from Co$_2$ to Co$_7$ is indeed approximately 0.7~\textmu$_\mathrm{B}$/atom, which is exactly what is expected from extrapolations from experimental data. Furthermore, one can conclude that LDA+DMFT already in its most simplified form (the Hubbard-I approximation) provides very accurate orbital moments except for Co$_2$, i.e. when hybridization effects are expected to be large. In this case, and in general for all systems with large hybridization effects, the more accurate exact diagonalization version of the LDA+DMFT method should be employed.  

From the discussion above it is clear that the calculated orbital moment for Co$_8$ and Co$_9$ within the \mbox{Hubbard-I} approximation is in good agreement with experiment. Then, from a comparison between LDA+DMFT calculations with and without hybridization effects included, we could show that the orbital moment for Co$_2$ to Co$_7$ is also approximately 0.7~\textmu$_\mathrm{B}$/atom. It is important to stress that this in principle only holds for neutral clusters. Future (XMCD) experiments should show whether this also holds for charged clusters. Very recently XMCD experiments have been performed on Co$_2^+$, for which the ground state is found to be of $2S+1=6$ and $L=1$ type~\cite{lauco2}. This result is obtained from a discussion, which is entirely based on the ratio of the spin and orbital moment. In this way difficulties due to the unknown number of d-holes, ion temperatue and degree of circular polarization due to the incident photon beam are circumvented. However, in this work also an estimation of the orbital and spin moment is made, i.e. respectively 0.29~\textmu$_\mathrm{B}$/atom and 1.18~\textmu$_\mathrm{B}$/atom. Both orbital and spin moment are thus found to be about a factor 2 smaller than what we obtain from our best LDA+DMFT calculations (Table~\ref{tabdmft}). Here we should note that the ratio of orbital and spin moment found by us, i.e. 0.24, is exactly the same as what was observed experimentally. Further, it is difficult to reconcile an orbital and spin moment of 0.29~\textmu$_\mathrm{B}$/atom and 1.18~\textmu$_\mathrm{B}$/atom for a ground state of $2S+1=6$ and $L=1$. Therefore, for completeness we also performed a LDA+DMFT calculation with hybridization effects included for Co$_2^+ $ (with the same inter atomic distance as used for Co$_2$). We find an orbital moment of 0.69~\textmu$_\mathrm{B}$/atom, which again shows the very small influence of the charge on the orbital moment. Furthermore, the ratio of orbital and spin moment is found to be 0.20, which is within their error bars, i.e. 0.24$\pm$0.04. 

Further, it is clear that a change of the Hubbard $U$ from 8 to 9~eV has little influence on the spin and orbital moments of Co$_{2}$ (Table~\ref{tabdmft}). The same holds for an increase of the interatomic distance from 2.2 to 2.4~\AA. Thus, although DMFT is supposed to work better for increasing cluster size due to the increasing number of nearest neighbors, the orbital moment of Co$_2$ is already in agreement with our expectation of 0.7~\textmu$_\mathrm{B}$/atom. However, note that according to an exact (many body) consideration, the sum of the spin and orbital angular moment along the dimer axis should be integer or half integer. From an inspection of Table~\ref{tabdmft} it is clear that this is not the case. This could be due to an overestimation of the spin, since it is subtantially larger than the values obtained by GGA and LDA+U (see Table~\ref{vaspso1}). Further, it is well known that approximate methods like GGA, LDA+U and LDA+DMFT can violate rigorous symmetry considerations.

\begin{figure}[!ht]
\begin{center}
\includegraphics[trim=100 25 30 20, clip, width=9cm, scale=0.5]{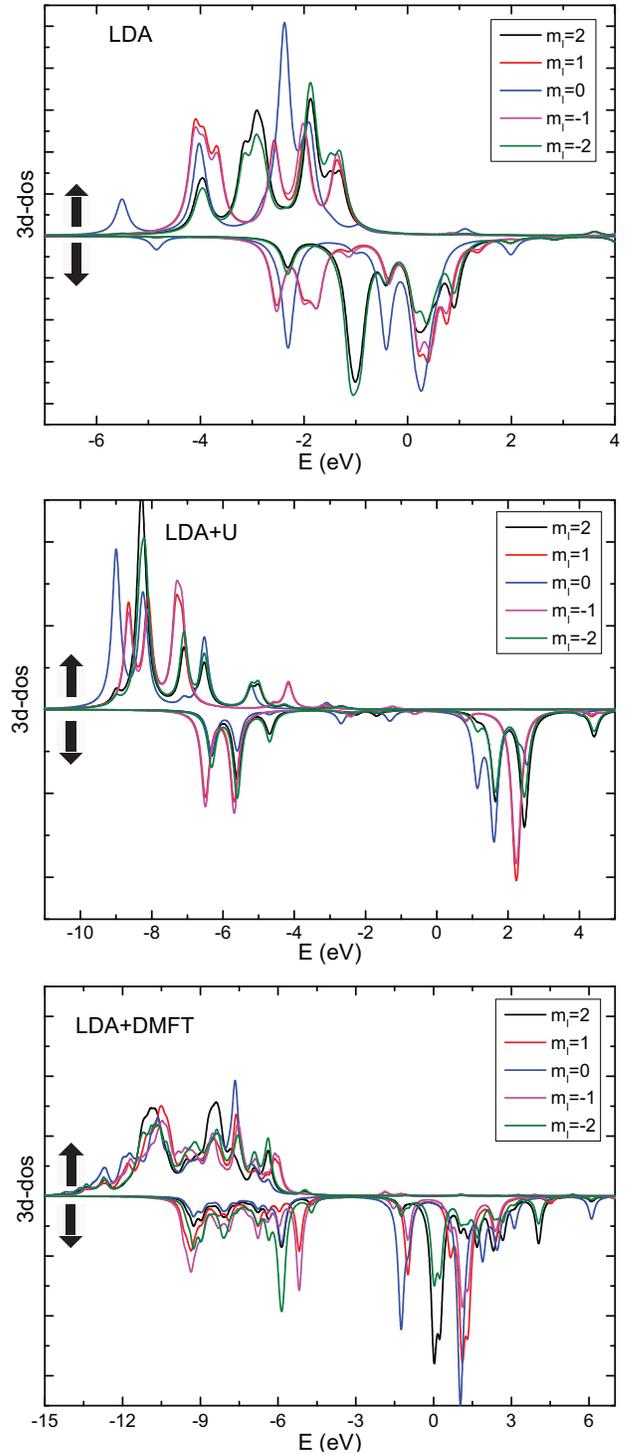}
\end{center}
\caption{The $m_{l}$-projected 3$d$ density of states for a planar Co$_{4}$ cluster with LDA (top), LDA+U (middle) LDA+DMFT with hybridization (bottom).} 
\label{dos}
\end{figure}

For Co$_{3}$ the LDA+DMFT calculations have been performed for two spin axes: one with a spin axis in the triangular plane and another with a spin axis orthogonal to the triangular plane, see Fig.~\ref{co234}. From Table~\ref{tabdmft} it can be seen that for the in-plane spin axis the orbital moment is 0.1~\textmu$_\mathrm{B}$/atom larger than for the out of plane spin axis. For both spin axes the orbital moment is in good agreement with the roughly expected orbital moment of 0.7~\textmu$_\mathrm{B}$/atom. The spin moment is a bit larger than obtained from GGA and LDA+U in Table~\ref{vaspso1}.

For Co$_{4}$ two different spin axes in the plane of the rhombus are considered (Fig.~\ref{co234}). As can be observed from Table~\ref{tabdmft}, the orbital moment is very similar for both spin axes. Further, both orbital moments are in good agreement with the 0.7~\textmu$_\mathrm{B}$/atom orbital moment, which is roughly expected. The spin moment is very similar to that obtained for GGA and LDA+U in Table~\ref{vaspso1}. 

\begin{figure}[!ht]
\begin{center}
\includegraphics[trim=50 25 30 20, clip, width=9cm, scale=0.5]{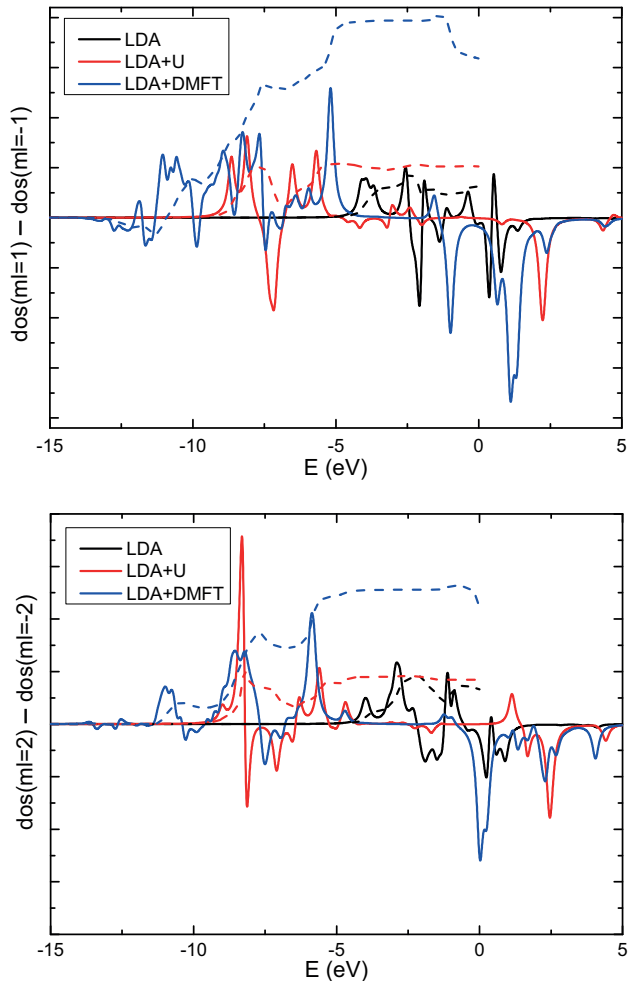}
\end{center}
\caption{For a planar Co$_{4}$ cluster the difference between the $m_{l}=1$ and $m_{l}=-1$ (top), and $m_{l}=2$ and $m_{l}=-2$ (bottom) density of states for LDA (solid black line), LDA+U(solid red line) and LDA+DMFT with hybridization (solid blue line). The dashed lines correspond to the integrals of these differences. } 
\label{diff_dos}
\end{figure}

In order to visualize the difference in orbital moment between the LDA, LDA+U and LDA+DMFT (with hybridization effects included) methods, we took the planar structure of Co$_{4}$ as a typical example to plot the projected 3$d$ density of states for (see Fig.~\ref{dos}). From this figure it can be observed that the density of states changes drastically between the methods. Furthermore, by a detailed inspection one may observe that the difference between the $m_{l}=1$ and $m_{l}=-1$, as wel as $m_{l}=2$ and $m_{l}=-2$ projected density of states increases when going from LDA to LDA+U and from LDA+U to LDA+DMFT. To see how this asymmetry carries over to the orbital magnetism, we plot these differences in Fig.~\ref{diff_dos}. Here the solid lines refer to the difference in the 3$d$ density of states of $m_{l}$ and $-m_{l}$, and the dashed lines correspond to the integrals of the these differences. From these dashed lines it is clear where and how the difference in orbital moment between the different methods occurs. In fact the enhanced orbital moment of the LDA+DMFT method, compared to the other two methods, is not the result of a $m_{l}$ projection or states in a narrow energy interval. Instead Fig.~\ref{diff_dos} shows that the LDA+DMFT calculations result in large contributions of the orbital magnetism over the entire occupied energy interval and for all $m_{l}$ projections (except $m_{l}=0$).

Finally, we would like to come back to Table~\ref{tabhia}. It would be interesting to see for what cluster sizes the orbital moment reaches the (hcp) bulk value of 0.13~\textmu$_\mathrm{B}$/atom~\cite{orbexp}. We speculate that a central atom in a cluster with nearest and next-nearest neighboring atoms will have a bulk like orbital moment. This speculation is based on the observation that for surfaces in general the third layer already behaves bulk like~\cite{surfbulk}. Thus, we expect that in order to obtain a bulk like total orbital moment the major part of atoms in a cluster should have nearest and next-nearest neighboring atoms.  



\section{Conclusion}
The size and direction of spin and orbital moments are determined by several interactions of a material, e.g. kinematic effects, crystal field interaction and on-site Coulomb repulsion between electrons. All this can amount to a complex dynamical interaction which critically influences all properties, in particular magnetism. The present investigation is mainly focused on spin and orbital magnetism of clusters, where we find that the orbital magnetism behaves differently from the large value of the atomic limit as well as the reduced value of bulk and thin films. We here focus on clusters since they are excellent model systems that allow for an investigation where different interactions can have different relative importance, and hence allow for a means to elucidate the importance of different contributions. We consider several levels of theory to undertake this investigation, with increasing level of accuracy, e.g. GGA, orbital polarization correction, LDA+U and hybrid functionals. We found that none of these approximations resulted in calculated orbital moments that are in agreement with experiments. Only when one considers a description based on multiple Slater determinants, as in the LDA+DMFT method, the theory predicts the orbital moments is in accordance with experiment. Thus, for a proper treatment of the orbital moment it is absolutely crucial to take the onsite Coulomb correlations accurately into account in a dynamical fashion. Furthermore, from comparing LDA+DMFT calculations with and without hybridization effects, we can conclude that the static crystal field potential is the dominant quenching mechanism for the orbital moment except for Co$_2$, where hybridization effects are also very important. Since LDA+DMFT becomes exact in the limit of negligible hybridization, it is not surprising that it already works for small cluster sizes, Co$_3$ to Co$_9$. Our findings in this work are relevant not only for Co clusters, but have bearing also for isolated Co atoms on substrates, e.g. as reported in Refs.~\onlinecite{gambardella,rau} or as impurities~\cite{brewer}. These studies can be summarized as all showing large orbital moments (in the range of $\sim$ 0.8 to $\sim$ 1.5 $\mu_B$/atom) in experiment, which was not reproduced by first principles theory (on GGA or LDA level). In these works, as well as in previous investigations,~\cite{eriksson,bruno,andersson} the effects of reduced symmetry, correlation effects associated with narrow bands, and spin-orbit effects of ligand orbitals were discussed. However, a clear understanding of which effect dominates for specific systems was not obtained. The present investigation clearly points to the importance of electron correlation as a general cause of large orbital magnetism of narrow band systems.




  \subsection*{Acknowledgements}
 We acknowledge support from the Swedish Research Council (VR), eSSENCE, STANDUPP, and the Swedish National Allocations Committee (SNIC/SNAC). The Nederlandse Organisatie voor Wetenschappelijk Onderzoek (NWO) and SURFsara are acknowledged for the usage of the LISA supercomputer and their support. The calculations were also performed on resources provided by the Swedish National Infrastructure for Computing (SNIC) at the National Supercomputer Center (NSC), the High Performance Computing Center North (HPC2N) and the Uppsala Multidisciplinary Center for Advanced Computational Science (UPPMAX). O.E. also acknowledges support from the KAW foundation (projects 2013.0020 and 2012.0031). M.I.K. acknowledges a support by European ResearchCouncil (ERC) Grant No. 338957. L.P. also acknowledges Dr. Davide Ceresoli for his support in the usage of the Quantum Espresso code. Further, L.P. also acknowledges Dr. Soumendu Datta for discussions on the VASP calculations, and Dr. Gustav Bihlmayer and Dr. Timo Schena for discussions on FLEUR calculations.

\appendix*
\section{}
Below the computational details are given for each of the used methods. Since all codes are ${\mathbf k}$-space codes, a supercell approach was used, with a large empty space between clusters that were repeated in a periodic lattice. In practice a large unit cell of at least 14~\AA~dimensions was used to prevent the interaction between clusters of different unit cells. The only ${\mathbf k}$-point considered was the $\Gamma$ point, and all calculations included the spin-orbit coupling.

Before providing all the computational details it is important to say something about the geometry of the clusters. Every theoretical consideration about clusters requires the (ground state) geometry. Although the spin and orbital moments can be obtained experimentally, it is a real challenge to probe the geometry of the cluster. Bulk-like experimental techniques, i.e. those based on x-ray diffraction, cannot be employed for obtaining the geometry of isolated clusters in the gas phase due the diluteness of the gas. The geometry of the cluster with the lowest total energy is considered as the cluster geometry. To obtain this structure properly, the geometries are calculated with DFT for all possible spin and orbital magnetic moments, in other words, spin states and electronic configurations~\cite{datta,agui1,agui2,lopez,castro}. Another method in selecting the proper cluster geometry is to compare experimental vibrational spectra with those obtained theoretically for the different geometries~\cite{jeroen}. The second method is especially useful in case of doubt about the computational total energies, for example, when the total energies of two or more structures are very close. 
 
 \subsection{RSPt}
 RSPt software (http://fplmto-rspt.org/) is a full-potential linearized muffin-tin orbital method (FP-LMTO) developed by Wills et al.~\cite{rsptbook}. In the calculations presented here the space was divided in muffin-tin spheres whose radius was of 1.95~a.u., and an interstitial region. The main valence basis functions included 4$s$, 4$p$ and 3$d$ states, while 3$s$ and 3$p$ states were treated as pseudocore in a second energy set~\cite{rsptbook}. Three kinetic energy tails were used for the 4$s$ and 4$p$ states, with values -0.3, -2.8 and -1.6~Ry. For the plain DFT calculations the LDA (PW) functional was used.
 
 RSPt includes an implementation of the orbital polarization correction (OPC) as described in Refs.~\onlinecite{rsptbook,olle}. The main idea of this correction is to include an approximate description of the second Hund's rule into the DFT problem. From a multipolar decomposition it can be shown that the orbital polarization correction term is contained in the LDA+U method~\cite{opcldau}. For the orbital polarization calculations the GGA (PBE) functional was used.
 
 The RSPt code was also used to perform the LDA+DMFT calculations both with and without hybridization effects included, where for both problems the exact diagonalization solver is used.  For details on the implementation of this routine see Refs.~\onlinecite{oscar,patrick,dmft2}. The local orbitals used in LDA+DMFT were constructed by considering only the so-called ``head'' of the LMTOs, which correspond to the MT orbitals of Refs.~\onlinecite{rsptbook,alexei}. In the case where hybridization effects are included the number of auxiliary bath states per 3$d$ orbital (used in the exact diagonalization) is one, i.e. there are ten 3$d$ states and ten auxiliary bath states to consider in the many-body problem. The fully localized limit (FLL) was used as the double counting correction. For the LDA+DMFT calculations the LDA (PW) functional was used.  
 
 Since this code is a collinear spin code with fixed spin quantization axis, different spin quantization axes were considered. Furthermore, the calculations performed with RSPt are for fixed geometry.

\subsection{VASP}
 The Vienna ab-initio simulation package (VASP) is a DFT implementation based on a pseudopotential augmented-plane-wave method~\cite{vasp}. As a cut-off of the plane wave basis set a kinetic energy of 400~eV was used. The calculations were considered converged for changes of the total energy smaller than $10^{-7}$~eV between two consecutive iterations. The geometry was considered converged, when the forces on all atoms were smaller than 5~meV/\AA. For the LDA+U calculations we employed the rotationally invariant formulation of Lichtenstein et al.~\cite{ldaulich} and the GGA (PBE) functional. For the plain DFT calculations the GGA (PBE) and hybrid (B3LYP) functionals were considered. Since the geometry of Co clusters has been extensively investigated in Ref.~\onlinecite{datta}, we used these ground state geometries and magnetic structures as starting points. Further, the calculations were spin polarized with non-collinearity, and the spin-orbit coupling was also included. In order to avoid to get trapped in a local minimum of the magnetic structure, different starting directions of the spin quantization axis were considered.
 

\subsection{FLEUR and SPEX}
 The FLEUR code is based on DFT and is an implementation of the full-potential linearized augmented plane wave (FLAPW) method~\cite{fleur}. As a cutoff for the plane waves 3.6~Bohr$^{-1}$ was taken, while $l_{cut}=8$ was used for the angular momentum. Moreover, the GGA (PBE) functional was employed. Based on the DFT calculations, the SPEX code~\cite{spex} was used in combination with the WANNIER90 code~\cite{wann1,wann2} to perform cRPA calculations~\cite{ersoy} of the Hubbard $U$ and Hund exchange $J$ parameters. The WANNIER90 code is used for the construction of the maximally localized Wannier functions (MLWF). In the construction of the MLWF's six states per Co atom are included, i.e. five $d$ states and the valence $s$ state. Finally, the geometry was fixed in these calculations, corresponding to the optimized geometry obtained with VASP in a collinear spin-polarized scalar relativistic (without spin-orbit coupling) approximation in GGA (PBE), which were also obtained in Ref.~\onlinecite{datta}. 

\subsection{Quantum ESPRESSO}
 Quantum ESPRESSO is a DFT implementation based on a pseudopotential plane wave method~\cite{espresso}. This code was used to evaluate the interstitial region contribution to the total orbital moment~\cite{ceresol}. The interstitial region is defined as the region outside the spheres around the atomic sites. These spheres were constructed with a radius of  2.0~a.u. For the plane wave basis a kinetic energy cut-off of 90~Ry was used. Furthermore, the GGA (PBE) functional was used. These calculations were performed for fixed geometries, which were also obtained from the scalar relativistic GGA VASP calculations, i.e. the same geometries as for the FLEUR/SPEX calculations were used.



\end{document}